\documentclass[runningheads]{llncs}
\usepackage[pdftex]{graphicx}
\usepackage[pdftex]{color}
\usepackage{bbding}
\usepackage{comment}
\usepackage{ntheorem}
\usepackage{url}
\newtheorem{hyp}{Hypothesis}

\begin{document}
\title{Influence of Selective Exposure to Viewing Contents Diversity}  
%
%

\author{Kota Kakiuchi\inst{1}$^($\Envelope$^)$  \and
Fujio Toriumi\inst{1} \and
Masanori Takano\inst{2} \and \\
Kazuya Wada\inst{2} \and 
Ichiro Fukuda\inst{2}}
\authorrunning{K.Kakiuchi et al.}
%
\institute{The University of Tokyo, Tokyo, Japan\\
\email{kakiuchi@torilab.net}\\
\and
CyberAgent, Inc., Tokyo, Japan}
%
\maketitle              
\begin{abstract}
Personalization, including both self-selected and pre-selected, is inevitable when tremendous amounts of media content are available. Personalization, which is believed to cause people to consume fewer diverse contents, can lead to fragmentation and polarization in society. Therefore, it is important to investigate the diversity of consumed contents over time. 
In this paper, first, we propose a framework to measure and analyze how the diversity of the consumed contents of users changes over time. In our framework, we introduce a new metric to measure content diversity based on our redefinition of diversity. Then, we investigate the relationship between selective exposure and content diversity changes using our framework and examine what factors encourage people to consume contents that are more diverse. We find that people autonomously consume more diverse contents from a macro-perspective without an external influence, suggesting that people are less likely to be fragmented and polarized, although from a micro-perspective they consume limited contents. We also obtain evidence that users who consume highly ambiguous contents tend to increase the diversity of their consumed contents.  
 
\end{abstract}

\section{Introduction}
We are surrounded by more and more media content, such as newspapers, radio, TV, and various web contents. However, we obviously lack the time or the ability to consume a significant percentage of everything. It is also troublesome to find interesting content or content that we actually need. Therefore, we sometimes consciously consume personalized contents and at other times unconsciously do so. In other words, we depend on personalization to find preferable content. \par

Personalization is divided into two main types: self-selected and pre-selected ~\cite{zuiderveen2016should}. Self-selected personalization comes from selective exposure, which is defined as a tendency for media consumers to select like-minded content. Pre-selected personalization, which comes from recommender systems or social networks that provide content based on our interests, can expose people to narrower viewpoints and ensnare them inside filter bubbles or echo chambers~\cite{pariser2011filter,sunstein2002law,sunstein2007republic}. These personalizations are believed to restrict people's exposure to less and less diverse content. If true, such personalization can increase fragmentation and polarization~\cite{stroud2010polarization,webster2005beneath,lelkes2013hostile,levendusky2013partisan,taber2006motivated,leeper2014informational}, and those can cause negative impact in society. For example, in politics, fragmentation and polarization might harm the public sphere, weaken democracy~\cite{vicke2013free}, and make it harder for society to reach a consensus on significant issues. In economics, fragmentation and polarization affect a positive externality from shared consumption~\cite{katz1985network}. Consumers can discuss their shared contents or experience and influence others. However, a lack of shared consumption decreases this effect. \par

Therefore, through the effect of selective exposure or a filter bubble, we must investigate the following two critical questions:
\begin{itemize}
\item How does the diversity of the contents consumed by users change over time? 
\item What behavior features contribute to such changes?
\end{itemize}
 If the above questions are investigated, the current content diversity situation might be grasped and an appropriate approach could be adopted to change it. Unfortunately, quantitative research about such questions remains inadequate.\par
In this study, we propose a method to measure how the content diversity of users quantitatively changes over time. We also clarify the relationship between selective exposure and content diversity changes with the method, using data from AbemaTV, a video streaming service in Japan that provides a user experience that resembles TV. Since TV greatly impacts people, it is important to analyze user experiences that are similar to TV for understanding societal fragmentation and polarization. In our framework, first, we represent the contents as vectors from a bipartite networks of users and contents to quantitatively evaluate content diversity. Next, we set a basic framework to measure the content diversity changes of users based on a previous method~\cite{nguyen2014exploring}. Finally, we introduce a new metric for evaluating the diversity of content lists. A conventional metric~\cite{ziegler2005improving} is inappropriate because diversity's definition is not suitable for the context of this study. Therefore, we redefine diversity and explain the variety of the accessed contents of each user. Then we measure how the diversity of the consumed content changes over time, present its changes, and scrutinize them from different perspectives with our proposed metric.\par

In addition, we investigate what kind of contents increase the diversity of users. We focus on the genres included in each piece of content and assume that such contents (those that include multiple genres) broaden user horizons. In this paper, we show uncontradicted results of our assumption from data analysis. \par

\subsubsection{Contributions}
In this paper, we make the following contributions:
\begin{enumerate}
\item We construct a framework to measure and analyze how the consumed content diversity of users changes over time, including leveraging the feature learning of contents from a bipartite network of users and contents and a new metric that measures content diversity. 
\item We clarify how the consumed content diversity of autonomous users changes over time with little external influence.
\item We show that consuming contents across multiple genres impacts the increase of content diversity. 
\end{enumerate}

\section{Related Work}
Our paper is related to a framework that measures content diversity and the consequences of selective exposure. We briefly review these two topics in this section. \par

First, we refer to papers that introduce a method to investigate how consumed contents change over time because of personalization. To the best of our knowledge, scant research exists on this topic. Hosanagar et al.~\cite{hosanagar2013will} studied whether personalization is fragmenting the online population on iTunes as a recommender platform. They built networks of consumers whose edges represented the similarity between consumer purchases and measured how the network properties changed over time. They found that consumers have more in common due to the influence of recommendations. Nguyen et al.~\cite{nguyen2014exploring} proposed a set of methods to measure how recommender systems affect the diversity of consumed contents over time and investigated the effect of collaborative filtering on MovieLens data. They found that users who follow recommendations consume more diverse contents than users who do not, although the diversity of the consumed contents narrowed in both cases. \par

Our framework is based on the method in Nguyen et al.'s study, which used a tag genome~\cite{vig2012tag}, which are user-generated tags, to quantitatively represent each piece of content. However, this approach is not standard because such tags are usually unavailable. In addition, although they used the average pairwise distance~\cite{ziegler2005improving} as a metric to measure the diversity of a content list, that metric is not appropriate in this context because its underlying definition of diversity does not mean what we really have to evaluate. Therefore, in our framework, we get the vector-representations of contents from the bipartite networks of users and contents because they can be obtained easily and introduce a new metric to measure content diversity. \par

Next, we refer to papers that show how selective exposure quantitatively affects fragmentation or polarization. Webster~\cite{webster2005beneath} evaluated audience fragmentation on the Nielsen Television Index and found that it was more advanced than is generally recognized. Audience polarization was also identified. Stroud~\cite{stroud2010polarization} investigated the relationship between selective exposure and political polarization on data from the National Annenberg Election Survey. She found strong evidence that selective exposure leads to polarization. Levendusky~\cite{levendusky2013partisan} showed that people who watched like-minded programs have more polarized opinions. Lelkes et al.~\cite{lelkes2013hostile} demonstrated that access to cable news and broadband Internet facilitates like-minded media consumption and leads to higher levels of affective polarization. \par
As shown above, various studies show that selective exposure leads to fragmentation and polarization~\cite{stroud2010polarization,webster2005beneath,lelkes2013hostile,levendusky2013partisan,taber2006motivated,leeper2014informational}, although concern also exists about measurement bias~\cite{prior2013media}. However, to the best of our knowledge, few studies have measured how the diversity of the consumed contents of users changes over time due to the effect of selective exposure. Scrutinizing it reveals whether the risk of fragmentation and polarization increases or decrease over time.  

\section{Data \& Method} \label{sec:datamethod}

In this section, we describe our dataset and explain our method that measures how the diversity of the content consumed by users changes over time. In the explanation of our method, we get vector-representations of contents from a bipartite network of users and contents by a network embedding method and design a framework to measure the changes of the content diversity based on a previous method ~\cite{nguyen2014exploring}. Finally, we introduce a new metric to measure the diversity of content lists in the context of this study. 
\subsection{Dataset}
In this study, we use the data of AbemaTV\footnote{https://abema.tv/} (Fig. \ref{abemahome}), which is a video streaming service that resembles TV, provided by AbemaTV, Inc.\footnote{http://abematv.co.jp/}. We use these data for the following two reasons. One, they do not have recommender systems. In most recent web services, recommender systems provide users with personalized information and stimulate them to increase their use of the services. However, for measuring the change of content diversity over time from the perspective of human nature, users need to select contents autonomously without any influence from recommender systems. The other reason is that this video streaming service closely resembles TV; it has channels and broadcasts programs based on a schedule. Even though more and more people use SNSs or websites to consume contents, TV continues to significantly impact the lives of people~\cite{ofcom2017communications}. Investigating the effect of selective exposure on TV is important. \par


\begin{figure}[tbh]
\begin{center}
\includegraphics[width=0.55\textwidth]{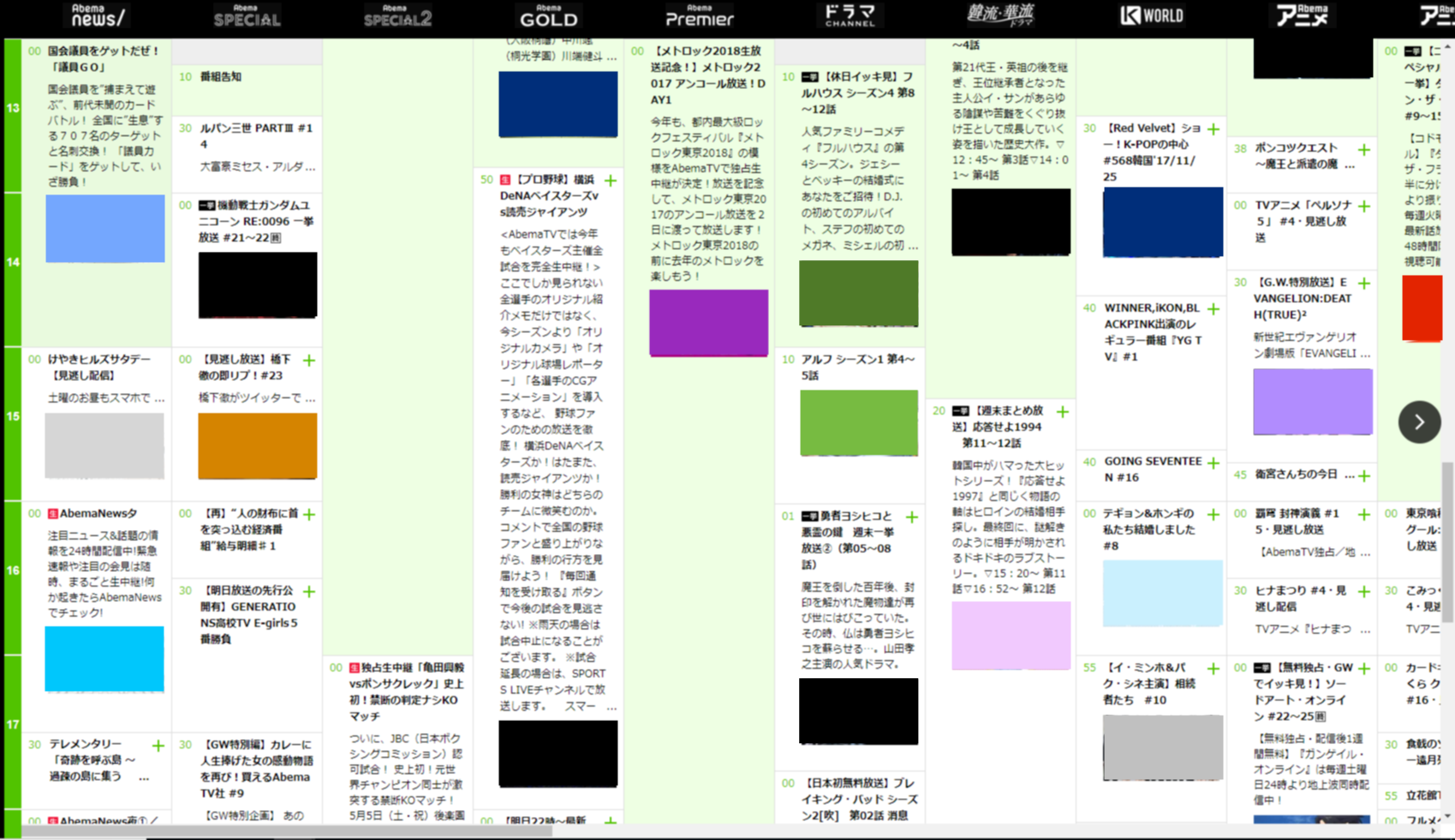}
\caption{A time schedule page of AbemaTV (An image of each program is masked because of portrait rights). Users autonomously select channels from this page or the main page.} \label{bipartite}
\end{center}
\end{figure}

We used the viewing history of the sampled users from June 1, 2017 to October 31, 2017. There are about 700,000 unique users and 13,000 programs every month in the data. \par
For measuring the changes of content diversity, we selected users who registered with AbemaTV on June 26, 2017 and July 2, 2017. On those two days, AbemaTV broadcast the shogi\footnote{Japanese Chess} matches of Sota Fujii, who is one of Japan’s best young professional shogi players. We selected these users because many joined the service under identical conditions in a short period. 45,556 users registered to watch these matches. \par

Additionally, we extracted users who watched the contents for more than 30 days during this five-month period and at least once in October 2017. In this study, we defined a program as having been watched if users viewed it for a minimum of five minutes. We identified 891 such users for our research.

\subsection{Distributed Representation of Contents}
To quantitatively measure the change of content diversity, each bit of content must be represented as a vector. We focus on the viewing behaviors of users to characterize each piece of content and employ the vector-representations of contents from the bipartite networks of users and contents by LINE (2nd)~\cite{line}, following a previous method \cite{pte}. \par

\begin{figure}[tbh]
\begin{center}
\includegraphics[width=0.9\textwidth]{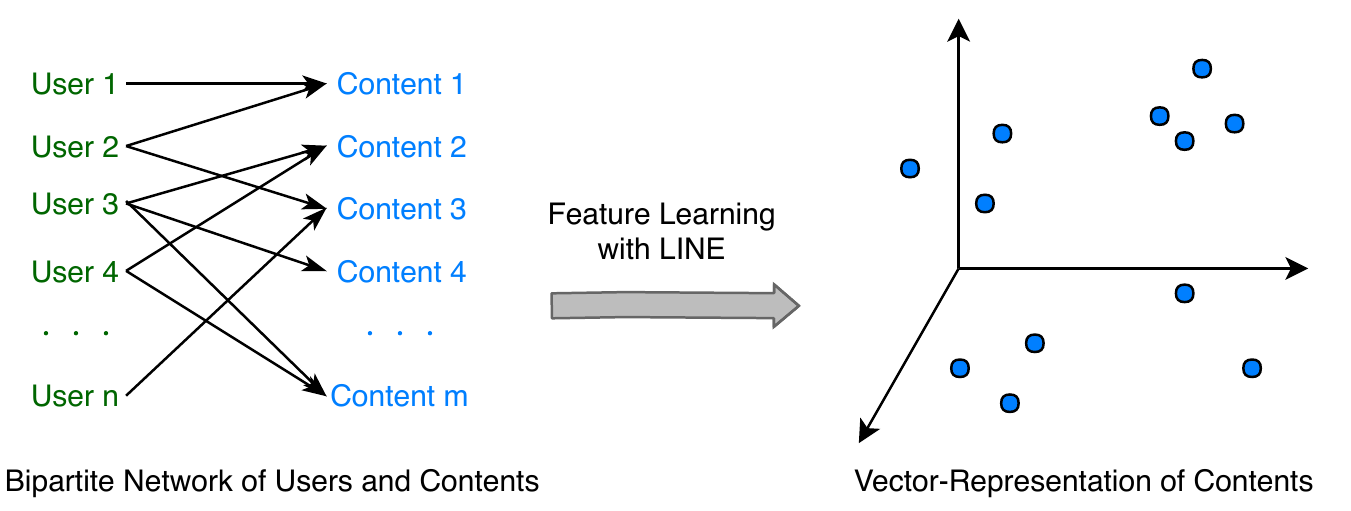}
\caption{Illustration that shows how to obtain vector-representations of contents} \label{bipartite}
\end{center}
\end{figure}

LINE (2nd) is a kind of network embedding method to assign nodes in a network to low-dimensional representations that preserve network structures effectively. LINE (2nd), which is based on the second-order proximity between vertices, assumes that similar vertices have similar neighbors. Therefore, vertices with high second-order proximity are embedded closely in a low-dimensional space. Consider bipartite network $G=(U\lor C, E)$, where U is a set of users, C is a set of contents, and E is a set of edges. The probability that user $u_i$ in U is connected to content $c_j$ in C is shown below: 
\begin{equation}
p(u_i|c_j) = \frac{exp(\overrightarrow{u_i}^{\mathrm{T}}\cdot \overrightarrow{c_j})}{\Sigma_{i'\in U} exp(\overrightarrow{u_{i'}}^{\mathrm{T}}\cdot \overrightarrow{c_j})},
\end{equation}
where $\overrightarrow{u_i}$ is the vector-representation of $u_i$ and $\overrightarrow{c_j}$ is the vector-representation of $c_j$. The empirical distribution of $p(u_i|c_j)$ is defined below:
\begin{equation}
\hat{p}(u_i|c_j) = \frac{w_{ij}}{d_j}, 
\end{equation}
where $w_{ij}$ means the weight of the edge between $u_i$ and $c_j$ (in this case, $w_{ij} = 0 \ or \ 1$) and $d_j$ means the out-degree of node $c_j$. To preserve the second-order proximity of the bipartite network, LINE (2nd) makes $p$ to be close to $\hat{p}$ by minimizing the KL divergence between them:
\begin{equation}
O = KL(\hat{p}, p).
\end{equation}
After omitting some constants, the objective function is represented:
\begin{equation}
O = - \sum_{(i, j) \in E} w_{ij}\log p(u_i | c_j)\label{O_without_ns}.
\end{equation}
Since Eq. (\ref{O_without_ns}) is too expensive to optimize, we used negative sampling~\cite{mikolov2013distributed} to write the objective function:
\begin{equation}
O = - \sum_{(i, j) \in E} \left\{ \log \sigma (\overrightarrow{u_i}^{\mathrm{T}} \cdot \overrightarrow{c_j}) + 
\sum_{k=1}^{K} E_{u_n \sim P_n(u)} \left[\log \sigma (\overrightarrow{u_n}^{\mathrm{T}} \cdot \overrightarrow{c_j})\right] \right\}\label{O_with_ns},
\end{equation}
where $\sigma$ is the sigmoid function, K is the number of negative edges, and $P_{n(u)} \propto (d_u)^{0.75}$ is the noise distribution shown in a previous work \cite{mikolov2013distributed}. \par
The objective function (\ref{O_with_ns}) is optimized with the stochastic gradient descent. Then we set the learning rate as $\rho_t = \rho_0 (1-\frac{t}{T})$, in which T is the number of edge samples and $\rho = 0.025$. The dimension of the embedding is set to 100. We set the number of negative samples as $K = 5$ and $T = 10$ billion. The bipartite network is constructed with the viewing history of users who watched more than 10 programs from June 1, 2017 to October 31, 2017. We got 20,878 embedding vectors of the contents. \par 
To confirm that those vectors properly represent the contents, we calculated the cosine similarity between the vectors and obtained the most similar programs to some programs. For example, when we look at {\sl Full House season 7, episode 10}, which is an episode of a famous American sitcom, the most similar programs are {\sl Full House season 7, episode 14}, {\sl Full House season 7, episode 11}, {\sl Full House season 7, episode 9}, {\sl Full House season 7, episode 13} and {\sl Full House season 7, episode 12} in descending order. When we look at {\sl FC Barcelona vs. SD Eibar}, which is a soccer game in Spain, the most similar programs are {\sl Athletic Bilbao vs. FC Barcelona}, {\sl FC Barcelona vs. Real Sociedad}, {\sl Real Sociedad vs. FC Barcelona}, a special program of FC Barcelona, and {\sl RCD Espanyol vs. FC Barcelona} in descending order. These examples show that the embedding vectors represent their contents well. 

\subsection{Framework to Measure Content Diversity Changes}\label{sec:measuring}
In this section, we design an experiment framework to measure the changes of content diversity based on a previously proposed method \cite{nguyen2014exploring}. 

\begin{figure}[tbh]
\begin{center}
\includegraphics[width=\textwidth]{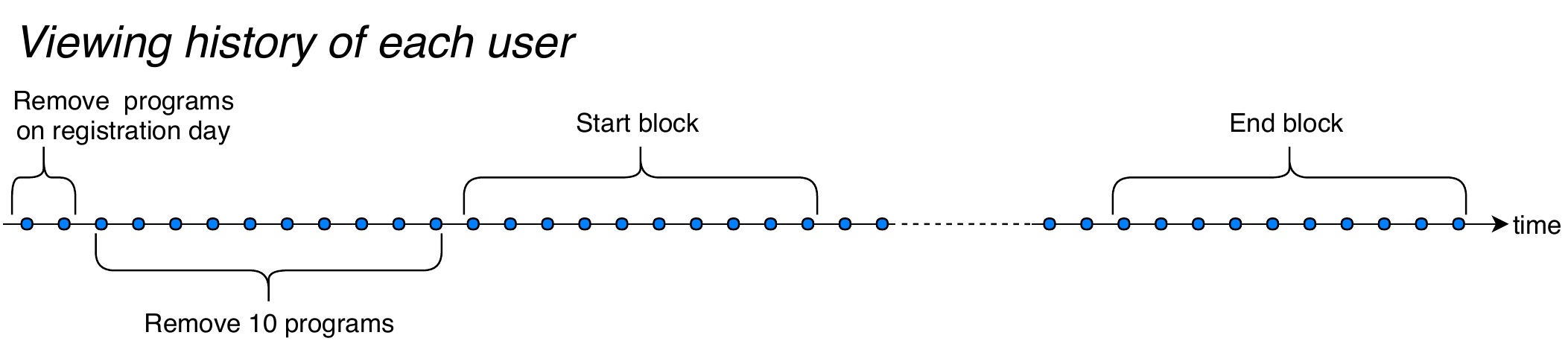}
\caption{Defining start and end blocks} \label{block}
\end{center}
\end{figure}

To measure the changes, we compared the first $n$ viewed programs and the last $n$ viewed programs of each user. We refer to the first $n$ viewed programs as a start block and the last $n$ viewed block as an end block. \par
Before defining a start block, we removed the viewed programs on the day the users registered and the first 10 programs because users need time to learn how to use AbemaTV. \par

Fig.~\ref{block} shows how to define a user's start and end blocks. We set $n = 10$ in this study. 
Then we calculated the content diversity of two blocks of each user by the metrics and made a content diversity distribution of the two blocks. The details of the metrics are explained in the next section. To investigate the change of the content diversity of users, we compared the content diversity distributions of the start and end blocks. 

\subsection{Metric of Content Diversity} \label{sec:metric}
In this section, we redefine diversity to evaluate the diversity of the consumed content and introduce a new metric named Cluster Diversity Entropy (CDE). 
\subsubsection{Definition of Diversity}
We assume that the diversity of the contents reflects the categories to which they belong. Such diversity depends on the perspective of how the contents are divided into categories. For example, consider political news, which is divided by such patterns as right-wing and left-wing, political parties, current political issues, and so on. When elections are held, the news should be divided from the perspective of political parties. In such a situation, we assume that the knowledge of people is more diverse when they are exposed to news about many political parties than when they are just exposed to information about one political party, even if it includes different topics. In other words, the experience of users is diverse when they are exposed to contents from plural categories under the assumption that the contents are divided into categories with appropriate granularity. \par

\subsubsection{Average Pairwise Distance}
A previous work~\cite{nguyen2014exploring} calculated the content diversity using the average pairwise distance~\cite{ziegler2005improving}, which is a general method to measure the diversity of a contents list. It is defined below:
\begin{equation}
D = \frac{\sum_{c_i \in L} \sum_{c_j \in L, c_j \neq c_i} d(c_i, c_j)}{|L|(|L| - 1)},
\end{equation}
where $c_i$ is each content and $L$ is the list of the contents. Since the vector-representations of the contents are sparse, we define distance function $d$ below: 
\begin{equation}
d(c_i, c_j) = 1 - \frac{c_i \cdot c_j}{|c_i||c_j|} \label{cosine}.
\end{equation}

However, is this metric actually appropriate to measure content diversity in the context of our study? How can we define what is a diverse experience for users? Indeed, the average pairwise distance implies the diversity of content lists, but it considers all the distances among the contents, including distances that don't really need to be considered. Therefore we propose a new metric called Cluster Diversity Entropy (CDE) to measure content diversity.

\subsubsection{Cluster Diversity Entropy}
Based on the above definition of diversity, we propose Cluster Diverse Entropy (CDE) as a metric to evaluate content diversity. CDE is calculated as follows: 

\begin{enumerate}
\item Apply a clustering method to all the contents and divide them into clusters.
\item Aggregate the clusters of the consumed contents.
\item Calculate the aggregation's entropy. 
\end{enumerate}

\begin{figure}[tbh]
\begin{center}
\includegraphics[width=\textwidth]{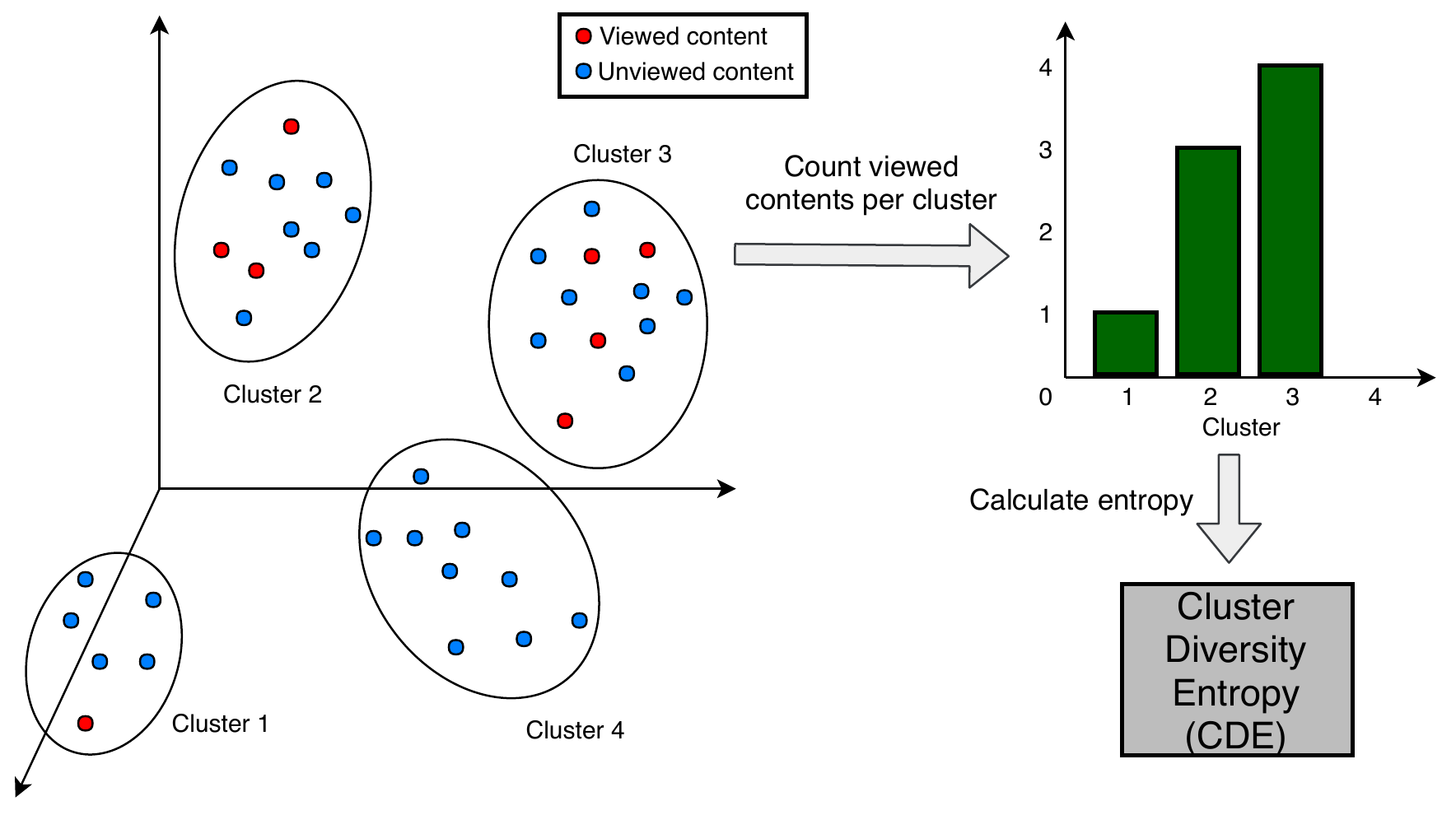}
\caption{Illustration of procedure to calculate CDE} \label{cluster_entropy}
\end{center}
\end{figure}

Fig.~\ref{cluster_entropy} illustrates the procedure to calculate CDE. First, we apply hierarchical clustering to vector-representations of 20,787 programs broadcast between June 1, 2017 and October 31, 2017. We leveraged Eq. (\ref{cosine}) as a distance function and an average linkage method at the point of the merging clusters. We adopted hierarchical clustering because it provides consistent results when the number of clusters $K_d$ are changed and comparing them is easy. In this experiment, we set $K_d=20$ based on the number of AbemaTV categories.\par

Then we extract clusters to which the viewing programs belong and count them as start and end blocks. We calculated the entropy of each block and finally got the content diversity distributions of both the start and end blocks. In this study we used both the average pairwise distance and CDE to compare the results. 

\section{Content Diversity Analysis} 
We present how the content diversity of users is changed by the effect of selective exposure using the method proposed in Section~\ref{sec:datamethod}. We show the result of both the average pairwise distance and CDE, and analyze them in detail. 

\subsection{Content Diversity Analysis using Average Pairwise Distance\label{sec:pairwise}}

\begin{figure}[tbh]
\begin{center}
\includegraphics[width=0.9\textwidth]{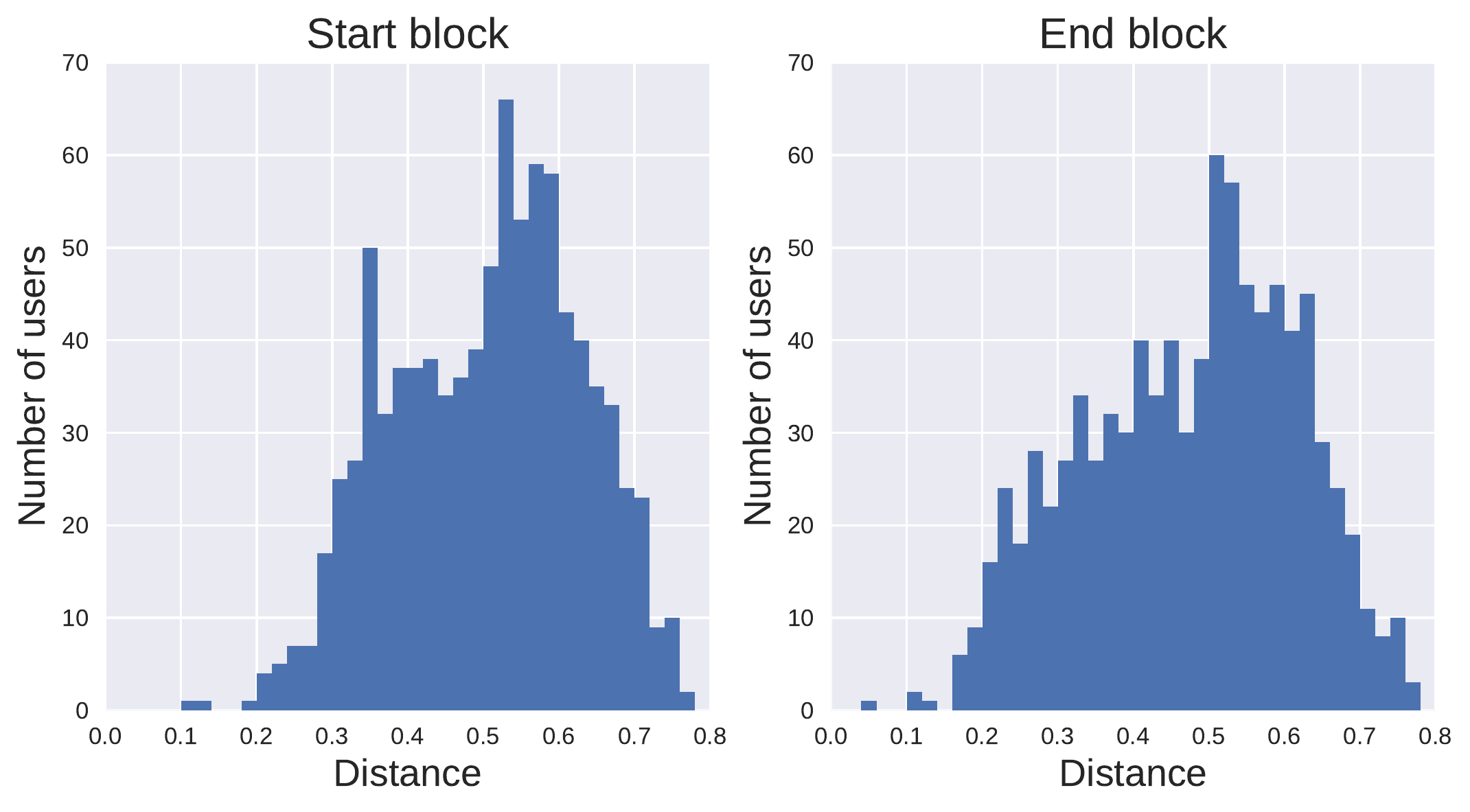}
\caption{Histogram of content diversity of start and end blocks, where x-axis is average pairwise distance of the blocks and y-axis is number of users.} \label{pairwise}
\end{center}
\end{figure}

To begin, we calculated the content diversity distributions of both the start and end blocks with the average pairwise distance. Fig. \ref{pairwise} shows their content diversity distributions.\par
To investigate the change of the content diversity of users, we measured the shift in the means of the content diversity distributions of the start and end blocks. We use a paired t-test and set the significance level to $0.01$.

Table \ref{diversity_change_table} shows the means of the content diversity distributions of the start and end blocks and the p-value for the t-test. Content diversity decreased over time from $0.5058$ to $0.4745$ when the average pairwise distance was used as a metric. Since the p-value is less than $0.01$, the difference of the two distributions is statistically significant. Hence the content diversity of users decreases over time. This coincides with a previous result~\cite{nguyen2014exploring}.

\begin{table}[h]
\begin{center}
\caption{Content diversity change between start and end blocks}\label{diversity_change_table}
\begin{tabular}{|c||c|c||c|}
\hline
 Metric & Start block & End block & p-value\\
\hline
Average pairwise distance &  0.5058 & 0.4745 & 3.59e-12\\
\hline
Cluster diversity entropy ($K_d=20$) & 0.3807 & 0.4357 & 2.81e-04\\
\hline
\end{tabular}
\end{center}
\end{table}

\subsection{Content Diversity Analysis using Cluster Diversity Entropy} \label{sec:CDE}

\begin{figure}[h]
\begin{center}
\includegraphics[width=0.9\textwidth]{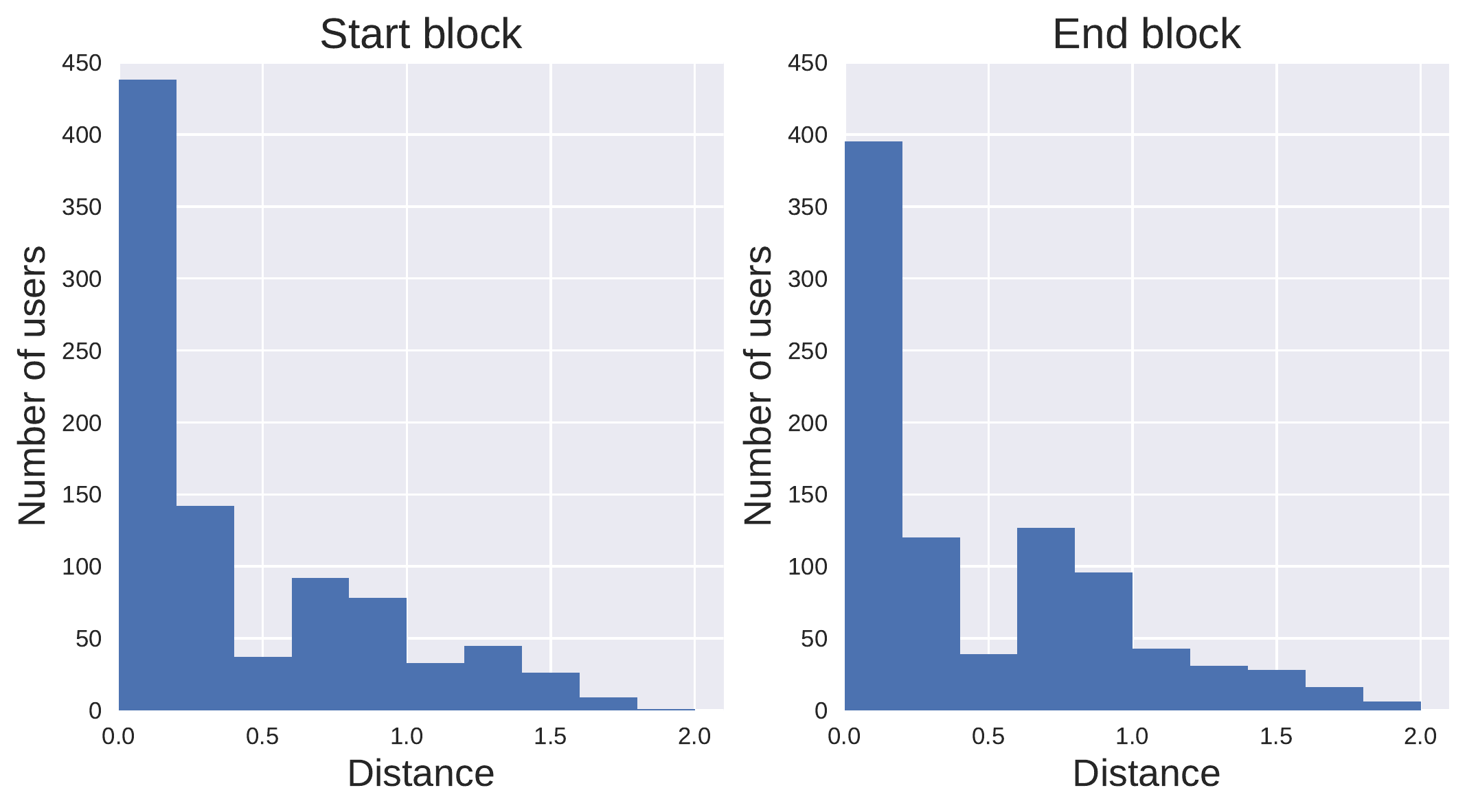}
\caption{Histogram of content diversity of start and end blocks, where x-axis is block’s CDE and y-axis is number of users.} \label{en_compare}
\end{center}
\end{figure}

Next, we calculated the content diversity distributions of both the start and end blocks with CDE. Fig.~\ref{en_compare} shows their content diversity distributions.
In the same way as in Section~\ref{sec:pairwise}, we measured the shift in the means of the content diversity distributions of the start and end blocks. \par

Table~\ref{diversity_change_table} shows the means of the content diversity distributions of the start and end blocks and the p-value for the t-test. The content diversity increased over time from 0.3807 to 0.4357 when CDE was used as a metric. Since the p-value is less than $0.01$, the difference of the two distributions is statistically significant. Hence, the content diversity of the users increases over time, which is contrary to the results in Section~\ref{sec:pairwise} and a previous work~\cite{nguyen2014exploring}. We analyze the cause for this difference in the next section.

\subsection{Analysis of Content Diversity Change with CDE}
In Section~\ref{sec:CDE}, we declared that the content diversity of users tends to increase when we used CDE when the number of clusters $K_d = 20$. Next we investigate how the number of clusters $K_d$ affect the result of measuring the content diversity change. \par

\begin{figure}[!h]
\begin{center}
\includegraphics[width=0.68\textwidth]{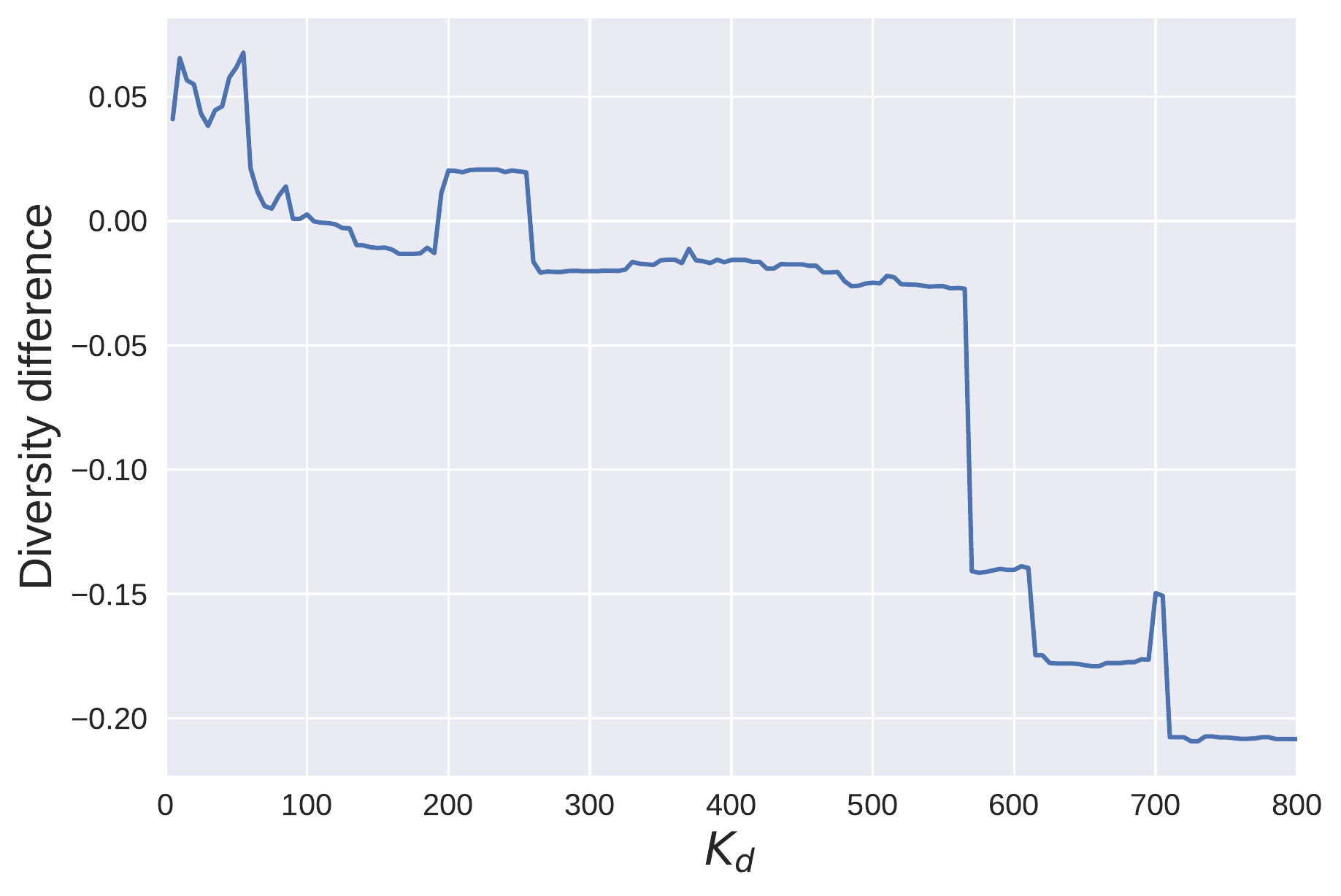}
\caption{Graph that shows how content diversity changes depending on number of clusters $K_d$, where x-axis is $K_d$ and y-axis is the difference between the mean of content diversity distribution of start and one end block.} \label{difference_change}
\end{center}
\end{figure}

We describe the relation between the number of clusters $K_d$ and the difference between the means of the content diversity distributions of the start block and the end block in Fig.~\ref{difference_change}. Here, $K_d$ corresponds to the perspective from which we view the contents. For example, in TV, when we only focus on large categories of contents (macro-perspective), $K_d$ is set at a small value. On the other hand, when we focus on every program of a series, its cast, or region (micro-perspective), $K_d$ is set to a larger value. Fig.~\ref{difference_change} shows that when $K_d$ is small, the content diversity of users is calculated at a higher value at the end block than at the start block. However, as $K_d$ grows, the diversity difference gets smaller and smaller. In other words, users gradually select fewer diverse contents over time from a micro-perspective, but they select more from a macro-perspective. This means that users gradually select narrow contents in a particular area but simultaneously select contents from different areas. \par
In Section~\ref{sec:pairwise}, we observed a decrease in the content diversity over time with the average pairwise distance. We suggest that this is because it reflects the micro-perspective more strongly. Since the average pairwise distance considers all the distances among the contents, when macro-level diversity is strongly considered with this study, we can evaluate content diversity more appropriately with CDE. \par
In this section, we clarified that the diversity of the consumed contents tends to increase over time when CDE is used. Therefore, users consume more diverse contents and are less likely to be fragmented or polarized when we only consider the effect of selective exposure without such external factors as recommender systems, social networks, or large events in the real world.

\section{Why do people broaden their horizons?} \label{horizon}
In Section~\ref{sec:metric}, we found that content diversity tends to increase over time when it is evaluated from a macro-perspective and argued that it should be measured from a macro-perspective in most cases. However, when we focus on each user, one user watches broader contents and another watches narrower contents. In this section, we investigate the factors that encourage users to watch contents with broader diversity. \par

We pose the following hypothesis to tackle this research question:

\begin{hyp}
Users who consume highly ambiguous contents tend to broaden their horizons. \label{hyp}
\end{hyp}
In this hypothesis, we define ambiguity of contents as how multiple categories they contain. Examples of highly ambiguous contents include a talk show in which some commentators discuss various topics (sports, politics or gossips) and a news in which we can have exposure to opinions of multiple political parties. We made this hypothesis because it is expected that highly ambiguous contents provide opportunities to have interests or to understand opinions in other areas.
We verified the hypothesis by the following experiment.

\subsubsection{Defining ambiguity of contents}
To quantitatively define the ambiguity of contents, we leverage the fuzzy c-means~\cite{Bezdek:1981:PRF:539444}, which is a soft clustering method. We set the fuzziness index to $m = 1.15$ and number of clusters $K_a = 20$, which is the same as $K_d$ in CDE. A fuzzy c-partitioned matrix was initialized based on the assumption that each program completely belongs to a CDE cluster. This initialization enables us to represent the contents based on the clustering results in CDE. \par

\begin{figure}[!h]
\begin{center}
\includegraphics[width=0.7\textwidth]{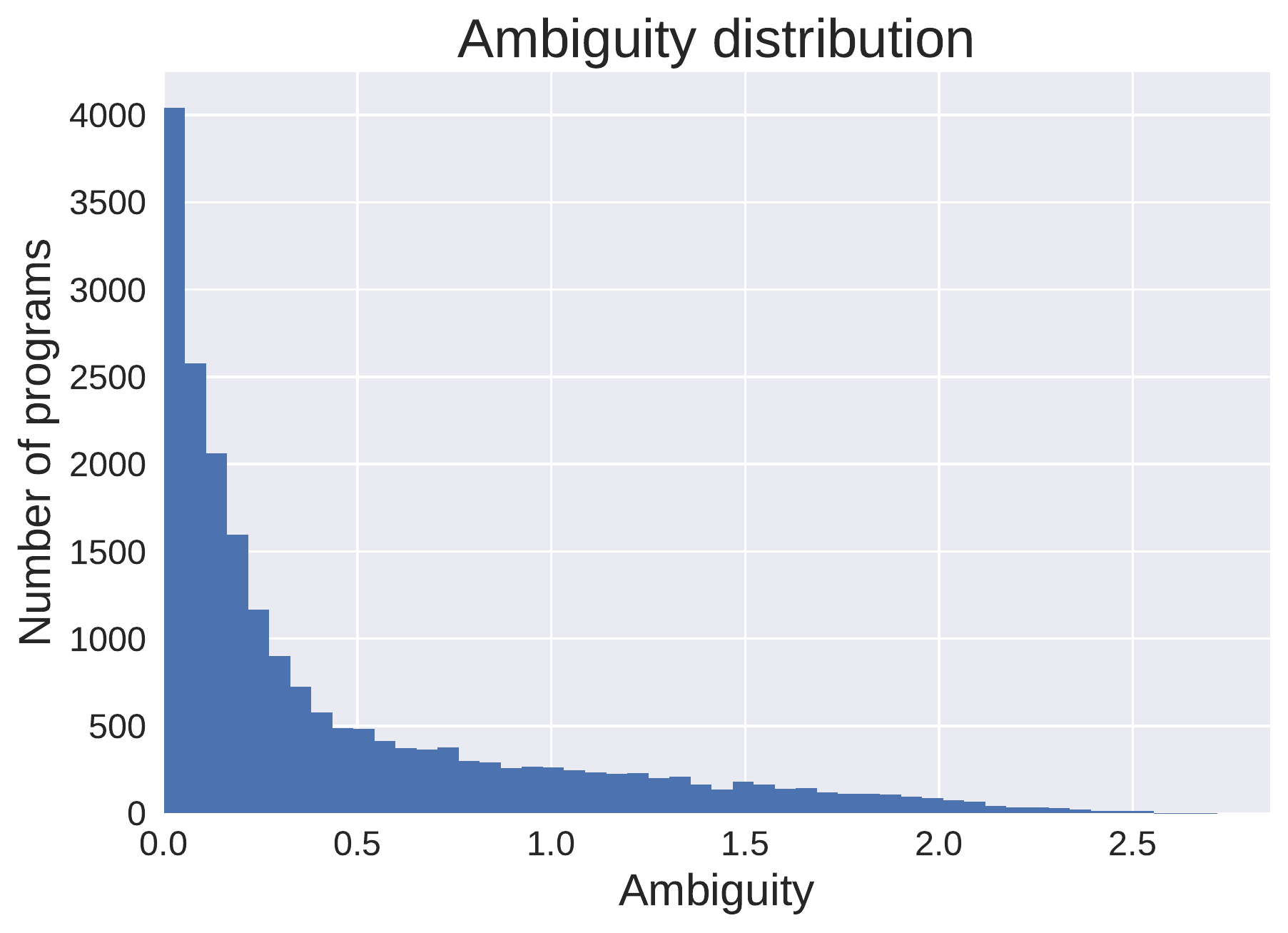}
\caption{Histogram of program ambiguity distribution, where x-axis is ambiguity of programs and y-axis is number of programs.} \label{ambiguity_distribution}
\end{center}
\end{figure}

We define the ambiguity of each bit of content as the entropy of the fuzzy c-means result which is a vector that represents the degree of the clusters to which it belongs. Fig.~\ref{ambiguity_distribution} shows the ambiguity distribution of all the programs and indicates that the ambiguity of most contents are near zero, but some exist across multiple genres.

\subsubsection{Experimental Settings}
Next we explain our experimental setting to verify our hypothesis. First, we extracted the maximum ambiguity of the viewing contents until the end block (maximum ambiguity) for each user. Then we made two groups: one that contains the top 300 max. ambiguity users (high ambiguity group) and another that contains the bottom 300 max. ambiguity users (low ambiguity group). We compared the increase of the content diversity from the start to end blocks between the two groups with a Welch's t-test. A p-value less than 0.01 is considered statistically significant.

\subsubsection{Experiment Result} 
We compared the increase of the content diversity of the high and low ambiguity groups. 

Table~\ref{ambiguity_table} shows the mean values of the content diversity increase of the high and low ambiguity groups. The increase is $0.1308$ in the former and $0.0218$ in the latter. Since the p-value is less than $0.01$, a null hypothesis, in which two groups have no statistical difference, is rejected at a significance level of $0.01$. The high ambiguity group significantly increases the diversity of the consumed contents more than the low ambiguity group. \par
Our experiment result verified our hypothesis. Perhaps contents with high ambiguity provide opportunities to users to learn more about other areas and broaden their horizons, although more detailed studies are necessary to support this assertion. 

\begin{table}
\begin{center}
\caption{Comparison of content diversity increase from start to end blocks between increase and constant groups}\label{ambiguity_table}
\begin{tabular}{|c||c|c||c|}
\hline
  & High ambiguity group & Low ambiguity group & p-value\\
\hline
\begin{tabular}{c}
\hspace{-2mm}Increase of\hspace{-8.5mm}\\ \hspace{-3.5mm}content diversity \hspace{-5.5mm}
\end{tabular}
&  0.1308 & 0.0218 & 5.98e-3\\
\hline
\end{tabular}
\end{center}
\end{table}

\section{Discussion}
In this paper, we propose a method that measures how the content diversity of users changes over time and investigate how content diversity changes by the effect of selective exposure with our method. We found that the content diversity of users tends to increase over time when it is observed with appropriate granularity. In addition, we analyzed the change in the details and found that content diversity generally increases from a macro-perspective and decreases from a micro-perspective. These results indicate that people tend to consume more diverse contents over time when we only consider the effect of selective exposure. This suggests that people are less likely to be fragmented and polarized over time if they are not affected by such external factors as web algorithms, social networks, or large events in the real world. \par

We also investigated what factors cause the content diversity of users to increase using the above result and found that the contents across multiple genres influence its increase. We believe that this factor effectively encourages people to consume more diverse contents. \par
In this study, we used our framework to investigate the changes of content diversity due to the effect of selective exposure on a video streaming service that resembles TV. Our proposed method, which is not specific to video streaming services, can be used to investigate various problems that are concerned about the content diversity caused by the effect of selective exposure, recommender systems, or social networks in various areas such as politics, economics or healthcare. \par 
We only considered the ambiguity of contents as a factor related to the increase of content diversity. However, perhaps other factors must be verified that contribute to the increases of content diversity with our method. If such knowledge is sufficiently stored, the consumed content diversity of users can be controlled to avoid fragmentation and polarization. \par
Our study suffers from some limitations. Even though we used data for a period of five months, which is relatively long, we must investigate with a longer period. Moreover, we assumed that AbemaTV users are not affected by such external factors as recommender systems or social networks, but they might be influenced by their friends in SNSs or in the real world.

%
%
%
%

\bibliography{reference}

\begin{thebibliography}{10}

\bibitem{Bezdek:1981:PRF:539444}
James~C. Bezdek.
\newblock {\em Pattern Recognition with Fuzzy Objective Function Algorithms}.
\newblock Kluwer Academic Publishers, Norwell, MA, USA, 1981.

\bibitem{hosanagar2013will}
Kartik Hosanagar, Daniel Fleder, Dokyun Lee, and Andreas Buja.
\newblock Will the global village fracture into tribes? recommender systems and
  their effects on consumer fragmentation.
\newblock {\em Management Science}, 60(4):805--823, 2013.

\bibitem{katz1985network}
Michael~L Katz and Carl Shapiro.
\newblock Network externalities, competition, and compatibility.
\newblock {\em The American economic review}, 75(3):424--440, 1985.

\bibitem{leeper2014informational}
Thomas~J Leeper.
\newblock The informational basis for mass polarization.
\newblock {\em Public Opinion Quarterly}, 78(1):27--46, 2014.

\bibitem{lelkes2013hostile}
Yphtach Lelkes, Shanto Iyengar, and Gaurav Sood.
\newblock The hostile audience: Selective exposure to partisan sources and
  affective polarization.
\newblock Technical report, Working Paper. Stanford, CA: Stanford University,
  2013.

\bibitem{levendusky2013partisan}
Matthew Levendusky.
\newblock {\em How partisan media polarize America}.
\newblock University of Chicago Press, 2013.

\bibitem{mikolov2013distributed}
Tomas Mikolov, Ilya Sutskever, Kai Chen, Greg~S Corrado, and Jeff Dean.
\newblock Distributed representations of words and phrases and their
  compositionality.
\newblock In {\em Advances in neural information processing systems}, pages
  3111--3119, 2013.

\bibitem{nguyen2014exploring}
Tien~T Nguyen, Pik-Mai Hui, F~Maxwell Harper, Loren Terveen, and Joseph~A
  Konstan.
\newblock Exploring the filter bubble: the effect of using recommender systems
  on content diversity.
\newblock In {\em Proceedings of the 23rd international conference on World
  wide web}, pages 677--686. ACM, 2014.

\bibitem{ofcom2017communications}
Ofcom.
\newblock International communications market report 2017.
\newblock available at:
  \url{https://www.ofcom.org.uk/__data/assets/pdf_file/0032/108896/icmr-2017.pdf},
  2017.
\newblock Accessed: April 30 2018.

\bibitem{pariser2011filter}
Eli Pariser.
\newblock {\em The filter bubble: What the Internet is hiding from you}.
\newblock Penguin UK, 2011.

\bibitem{prior2013media}
Markus Prior.
\newblock Media and political polarization.
\newblock {\em Annual Review of Political Science}, 16:101--127, 2013.

\bibitem{stroud2010polarization}
Natalie~Jomini Stroud.
\newblock Polarization and partisan selective exposure.
\newblock {\em Journal of communication}, 60(3):556--576, 2010.

\bibitem{sunstein2002law}
Cass~R Sunstein.
\newblock The law of group polarization.
\newblock {\em Journal of political philosophy}, 10(2):175--195, 2002.

\bibitem{sunstein2007republic}
Cass~R Sunstein.
\newblock Republic. com 2.0.
\newblock 2007.

\bibitem{taber2006motivated}
Charles~S Taber and Milton Lodge.
\newblock Motivated skepticism in the evaluation of political beliefs.
\newblock {\em American Journal of Political Science}, 50(3):755--769, 2006.

\bibitem{pte}
Jian Tang, Meng Qu, and Qiaozhu Mei.
\newblock Pte: Predictive text embedding through large-scale heterogeneous text
  networks.
\newblock In {\em Proceedings of the 21th ACM SIGKDD International Conference
  on Knowledge Discovery and Data Mining}, pages 1165--1174. ACM, 2015.

\bibitem{line}
Jian Tang, Meng Qu, Mingzhe Wang, Ming Zhang, Jun Yan, and Qiaozhu Mei.
\newblock Line: Large-scale information network embedding.
\newblock In {\em Proceedings of the 24th International Conference on World
  Wide Web}, pages 1067--1077. International World Wide Web Conferences
  Steering Committee, 2015.

\bibitem{vig2012tag}
Jesse Vig, Shilad Sen, and John Riedl.
\newblock The tag genome: Encoding community knowledge to support novel
  interaction.
\newblock {\em ACM Transactions on Interactive Intelligent Systems (TiiS)},
  2(3):13, 2012.

\bibitem{vicke2013free}
Vaira V{\=\i}{\c{k}}e-Freiberga, Herta D{\"a}ubler-Gmelin, Ben Hammersley, and
  Lu{\'\i}s Miguel Poiares~Pessoa Maduro.
\newblock A free and pluralistic media to sustain european democracy.
\newblock 2013.

\bibitem{webster2005beneath}
James~G Webster.
\newblock Beneath the veneer of fragmentation: Television audience polarization
  in a multichannel world.
\newblock {\em Journal of Communication}, 55(2):366--382, 2005.

\bibitem{ziegler2005improving}
Cai-Nicolas Ziegler, Sean~M McNee, Joseph~A Konstan, and Georg Lausen.
\newblock Improving recommendation lists through topic diversification.
\newblock In {\em Proceedings of the 14th international conference on World
  Wide Web}, pages 22--32. ACM, 2005.

\bibitem{zuiderveen2016should}
Frederik Zuiderveen~Borgesius, Damian Trilling, Judith Moeller, Bal{\'a}zs
  Bod{\'o}, Claes~H de~Vreese, and Natali Helberger.
\newblock Should we worry about filter bubbles?
\newblock 2016.

\end{thebibliography}
\bibliographystyle{plain}

\end{document}